\documentclass[sigconf]{acmart}
\AtBeginDocument{%
  \providecommand\BibTeX{{%
    \normalfont B\kern-0.5em{\scshape i\kern-0.25em b}\kern-0.8em\TeX}}}

\setcopyright{acmcopyright}
\copyrightyear{2022}
\acmYear{2022}
\acmDOI{XXXXXXX.XXXXXXX}

\acmConference[ICAIF '22]{}{November 02--04,
  2022}{New York, NY, USA}
\acmPrice{15.00}
\acmISBN{978-1-4503-XXXX-X/18/06}

\newcounter{noteMCctr} 

\definecolor{colour3}{RGB}{178,55,250} 
\newcommand{\mc}[1]{\textcolor{black}{{{}}#1}}

\begin{document}

\title[Asset Classes' Behaviour in Returns-Driven Macro Regimes]{Returns-Driven Macro Regimes and Characteristic Lead-Lag Behaviour between Asset Classes}

\author{Deborah Miori}
\authornote{Work done as part of a collaboration with the Fixed Income team at Fidelity Investments Inc., FMR, London UK.}
\email{deborah.miori@maths.ox.ac.uk}
\affiliation{%
  \institution{University of Oxford}
  \streetaddress{Andrew Wiles Building, Woodstock Rd}
  \city{Oxford}
  \country{United Kingdom}
  \postcode{OX2 6GG}
}

\author{Mihai Cucuringu}
\email{mihai.cucuringu@stats.ox.ac.uk}
\affiliation{%
 \institution{University of Oxford}
 \streetaddress{24-29 St Giles’}
 \city{Oxford}
 \country{United Kingdom}
 \postcode{OX1 3LB}}

\renewcommand{\shortauthors}{Miori and Cucuringu}

\begin{abstract}
  We define data-driven macroeconomic regimes by clustering the relative performance in time of indices belonging to different asset classes.  We then investigate lead-lag relationships within the regimes identified. Our study unravels market features characteristic of different windows in time and leverages on this knowledge to highlight market trends or risks that can be informative with respect to recurrent market developments. The framework developed also lays the foundations for multiple possible extensions.
\end{abstract}




\maketitle

\section{Introduction}
The broader dynamics of financial markets can significantly vary across time.  These can define periods of different macroeconomic regimes, which are clustered moments of persistent market conditions that can be characterised by external macroeconomic trends.

The importance of regime identification mainly lies in its implications and impact on asset allocation and portfolio construction, as highlighted in \cite{regimes-asset-allocation}. This is a developing area in academic research, where one main branch of efforts concentrates on network analysis of correlation and causality matrices of multi-asset returns. A review of the related research is available in \cite{review_clustering_financial_markets}, while \cite{survey_clustering_dynamic_networks} is an extensive survey on community detection in dynamic networks. 
The relevance of correlation studies is motivated by the higher interconnection of stock returns characteristic of periods of market distress \cite{correlation-stress}. Indeed, \cite{subtract-market-mode} and \cite{stock-market-temporal-network} are examples of clustering studies on equities via community detection on the minimum spanning trees 
built from returns' correlation matrices.
The former paper studies data from different stock markets at a frequency from minutes to daily scale, while the latter constructs risk diversified portfolios from centrality measures on the correlation-based evolving network.

However, the literature lacks an extensive study of return correlations and causalities for securities belonging to \textit{multiple} asset classes. In this paper, we pursue a detailed investigation of the topic, fueled by its strong  potential to shed further light on macro regimes identification and characterisation. 


The rest of this paper is structured as follows. Section \ref{sec:data} proceeds with a description of the two data sets considered in our analysis. Section \ref{sec:meth} details the methodology adopted for correlation analyses and lead-lag clusters identification. Sections \ref{sec:resIdenReg} and  \ref{sec:resLeadLag}   discuss the results for the two considered tasks. Finally, Section \ref{sec:concl} concludes our work with some final remarks and future research directions.


\section{Data} \label{sec:data}

\subsection{Data set 1: Long-Term (LT) monthly indices}

The first data set considered is provided by Fidelity Investments Inc. This is an internal set of $33$ indices belonging to different asset classes, whose majority of levels have been reconstructed at the last day of each month, starting from  January $1921$. Due to some remaining missing entries, we decide to study the $23$ time series that show full data from February $1926$ to April $2022$. The related names are reported in Appendix \ref{A-LT-data}, where we assign to each index an identifier from $0$ to $22$ for ease of reference. As for the asset classes,  indices $0$ to $9$ relate to equity, $10$ is commodity, $11$ to $21$ are fixed income and $22$ acts as a proxy for cash. For each time series, we compute returns as percentage changes of the related levels.


\subsection{Data set 2: Bloomberg (BBG) daily indices}

To have a more resilient and holistic view of the full financial market, we download and pre-process daily levels of a broad set of indices belonging to different asset classes from Bloomberg (BBG). We aim to strike a balance between a too agglomerated or too granular view, paying attention to the final number of items 
in each class. Our data set consists of $231$ daily time series, 
$43$ of which belong to the class of commodities, $29$ to currencies, $67$ to equities, $29$ to bond spreads, $12$ to volatilities and $76$ to interest rates. The full list of indices for each class is reported in Appendix \ref{A-BBG-data}. The data start on $30$ September $2005$, and end on 1 July 2022, where only trading days are kept. Each time series is converted into returns by calculating percentage differences or simple differences
for interest rates.

\section{Methodology}  \label{sec:meth}
Our first aim is to identify a reliable division of time into financial regimes from a data-driven point of view, partly extending \cite{papenbrock} and \cite{correlation-changes-future-volatility}. Each regime is then characterised by summary statistics and relationships between the returns of different asset classes. While we consider both the LT and BBG data sets for this part of the analysis, the main focus will overall be on the latter, due to its broader range of asset classes available and daily frequency.

The second step of our analysis investigates lead-lag relationships between clusters of assets, for each uncovered regime. The results are then tested for profitability of an investment on the lagging assets.

\subsection{Empirical identification of regimes}
\label{method-regimes}

\paragraph{Correlation of asset returns.} For each sliding window of time $\Delta T$, we compute the Pearson's linear correlation $\rho_P$, Spearman monotonic correlation $\rho_S$ and Kendall rank correlation $\tau_K$ between the time series of indices' returns. Precisely, we consider 
\begin{equation}
        \rho_P = \frac{
        \sum_{t \in \Delta T} (x_t-\overline{x})(y_t-\overline{y})
        }{
        \sqrt{
        \sum_{t \in \Delta T} (x_t-\overline{x})^2
        \sum_{t \in \Delta T} (y_t-\overline{y})^2
        }
        },
\end{equation}
\begin{equation}
        \rho_S = 1 - \frac{6\sum_{t \in \Delta T} d_t^2}{l(l^2-1)},
\end{equation}
\begin{equation}
        \tau_K = \frac{(\textsc{no. concordant pairs}) - (\textsc{no. discordant pairs})}{(\textsc{no. pairs})},
\end{equation}
where $(x_t,y_t), t \in \Delta T$ are the joint observations for each pair of time series $X$ and $Y$, $\overline{x}, \overline{y}$ are the related averages, $d_t = \textsc{rank}(x_t) - \textsc{rank}(y_t)$ and $l$ is the length of $\Delta T$. Concordant or discordant observations are counted from the number of pairs of observations $(x_{t_1},y_{t_1})$ and $(x_{t_2},y_{t_2})$ whose sort order agrees or not. We aim for $\frac{l}{N}\sim 2$, where $N$ is the number of assets considered. In this way, we avoid too short time-windows that would introduce statistical noise and we are more likely to achieve reliable correlations \cite{how-long-is-enough}.

Since we are dealing with time series, we lower the importance of points further away in history by applying hyperbolic weighting to the Kendall correlation. Indeed, $\tau_K$ measures the ordinal association between each pair of time series and we can intuitively add weight $\frac{1}{r+1} + \frac{1}{s+1}$ for each exchange between elements with rank $r,s \geq 0$. Some of the benefits of weighted correlations are mentioned in  \cite{weighted-correlations}, while unfortunately there is 
no known distribution-based test that allows 
to infer a 
$p-value$ of this weighted Kendall correlation $\tau_K^w$.

\paragraph{Market similarity in time and regimes.} After obtaining a set of correlation matrices for different points in time, we compute the related pairwise similarities. We consider two relevant measures:
\begin{enumerate}
    \item Cophenetic correlation \cite{cophenetic}. We transform each correlation matrix $C$ into a distance matrix $D$ by computing $D=\sqrt{2\cdot(1-|C|)}$. Then, we pursue hierarchical clustering via the linkage algorithm with the average method. This implies that a node is added to the cluster to which there is smallest average distance in an iterative fashion. Finally, the cophenetic correlation is a measure of the correlation between the distances of points in the feature (Euclidean) space and on the dendrogram generated by the clustering.
    Therefore, we can employ the distance matrix and dendrogram of different points in time to measure the related similarity. We follow this idea and compute the cophenetic correlation in both directions (then averaged) for each pair of points in time.
    \item Metacorrelation. We flatten the matrices of assets' correlations and, for each pair, compute Pearson's correlation \cite{risk-persistence}.
\end{enumerate}
Next, we perform Principal Component Analysis (PCA) on the resultant matrix of similarity between points in time. This is done to extract the axes that explain the greatest variance between the data and project the latter on them. We cluster dates in this new space via KMeans++ to identify different financial regimes. The optimal number of groups/regimes is chosen following the \textit{elbow method}, i.e. looking for the rough optimal point with lowest inertia and lowest number of clusters on the related plot. Inertia measures the distance between each data point and its centroid, squares it, and sums these results across each one cluster. Finally, we need to check the stability of the results while perturbing the chosen parameters.

\paragraph{Network study of the evolution of correlations.} To increase the confidence on the identified regimes, we compare them with the results obtained following another approach. We start from our set of correlation matrices between assets returns at different points in time and retain only entries with magnitude above a threshold. From each matrix, we build the related signed network $G=(I,E)$. The set of nodes $I$  are indices and edges $(i, j) \in E$ with $i,j \in I$ have weights $w_{ij}$ from the non-zero correlations. We save the giant components and cluster each final graph via the symmetric SPONGE algorithm \cite{sponge}. This is a spectral algorithm that aims at having the maximum number of positive edges within clusters while having highest number of negative edges between clusters. To choose the number of communities to look for, we compute the signed modularity \cite{signed_modularity}
\begin{equation}
        Q_{signed} = Q_+ \cdot \frac{W_+}{W_++W_-} - Q_- \cdot \frac{W_-}{W_++W_-},
\label{signed-mod}
\end{equation}
where $W_{\pm}$ are total weights and $Q_{\pm}$ modularities of the subgraphs generated by the subset of positive or negative ($\pm$) edges. Modularity is computed following \cite{modularity} and measures the strength of division of a network into modules. The overall motivation behind Eq. \ref{signed-mod} is to measure the trade-off between the tendency of positive weights to form communities, and of negative weights to separate them.

Once we have a clustering for each chosen point in time, we compare communities at consequent periods and compute a measure of the related stability. This is achieved by calculating the Adjusted Rand Index (ARI), which counts pairs of indices that are assigned to the same or different clusters at distinct points in time. Whenever there is a significant drop in stability of the communities identified, we expect a regime change to have happened.

\paragraph{Characterisation of the identified regimes.} Once regimes have been identified, we proceed by studying their peculiar features. We compute the average correlation matrix for each regime by considering the points in time assigned to it. Then, we also compute the related average mean returns $ret$, average standard deviations of returns $\sigma_{ret}$ and annualised Sharpe Ratios $S=\frac{ret}{\sigma_{ret}} \cdot \sqrt{252}$ of the underlying asset classes. 

Finally, we build the summary network of correlations for each regime and the related distance graph. On the former, we run again the symmetric SPONGE algorithm to investigate the different relationships between clusters and their composition across asset classes. On the latter, we compute the betweenness centrality \cite{betweenness} to understand which indices have the highest influence over the flow of information or shock propagation in the graph of each regime.


\subsection{Lead-lag clusters specific to regimes}
\label{sec:ll}

\paragraph{Causality relations in regimes.} After having identified our macro regimes, we can investigate whether any latent significant lead-lag relationships could be extracted. Thus, we fragment our time series of indices' returns into the related regimes. For each regime, we compute the Granger causality \cite{granger} between each pair of indices for lags $g \in [\pm 1, \pm 5, \pm 10, \pm 16, \pm 21, \pm 42]$ trading days and retain  results at the $0.05$ significance level. 
In the rare scenario that both indices in a pair lead or lag each other, we subtract the weakest relationship from the strongest and keep the deflated result, \mc{essentially leading to a skew-symmetric lead-lag matrix}. The optimal lag $g^R$ for each regime $R$ is chosen as the one leading to the highest number of significant Granger causalities, since we are interested in an agglomerated informative view.

Each resultant causality matrix is used to build a directed network for the related regime, from which we can identify leading and lagging groups of indices via the Hermitian clustering algorithm in \cite{hermitian-clustering}. \mc{This algorithm aims to cluster nodes according to the similarity in their outgoing and incoming edge patterns, and was also shown to be effective at exposing pairs of clusters $(A,B)$ such that there is a high \textit{imbalance} in the directional flow of edges between $A$ and $B$. This means that the majority of the edges flow cluster $A$ to cluster $B$, and very few in the opposite direction. In light of our analysis, an imbalance of such type between $A$ and $B$ would translate into a lead-lag relationship between the times series in $A$ and those in $B$.}   
To define the number of communities to look for, we leverage on the idea that there is structural similarity between the different asset classes. These classes might lead or lag each other, and thus we favour a clustering that generates groups with highest inner coherence in composition. As a measure of goodness, we thus choose the $v-measure$ \cite{v-measure}
\begin{equation}
    v = (1 + \beta) \cdot \frac{\textsc{homogeneity} \cdot \textsc{completeness}} {(\beta \cdot \textsc{homogeneity} + \textsc{completeness})}.
\end{equation}
Here, $\beta$ weights our interest between focusing on homogeneity or completeness of the extracted clusters with respect to the related composition across asset classes.

\paragraph{Investment implications.} As a final step, we compare the performance of an investment based on lead-lag clusters specific to each regime against a vanilla benchmark that bets uniformly on all assets. For the former, we consider the most leading cluster in each regime $R$, compute the average return of the related subset of indices over the past $g^R$ days, and use its sign as a signal to buy or sell assets uniformly in the most lagging cluster. The positions are closed after $g^R$ trading days. The average returns between the two strategies are then compared, despite being generated by in-sample analyses. Indeed, our aim is simply to show further features characteristic of regimes, and lay the foundations for an investment framework that will be developed in future work.


\section{Results: Identification of regimes}    \label{sec:resIdenReg}

\subsection{LT data set}

\begin{figure}[h]
  \centering
  \includegraphics[width=\linewidth]{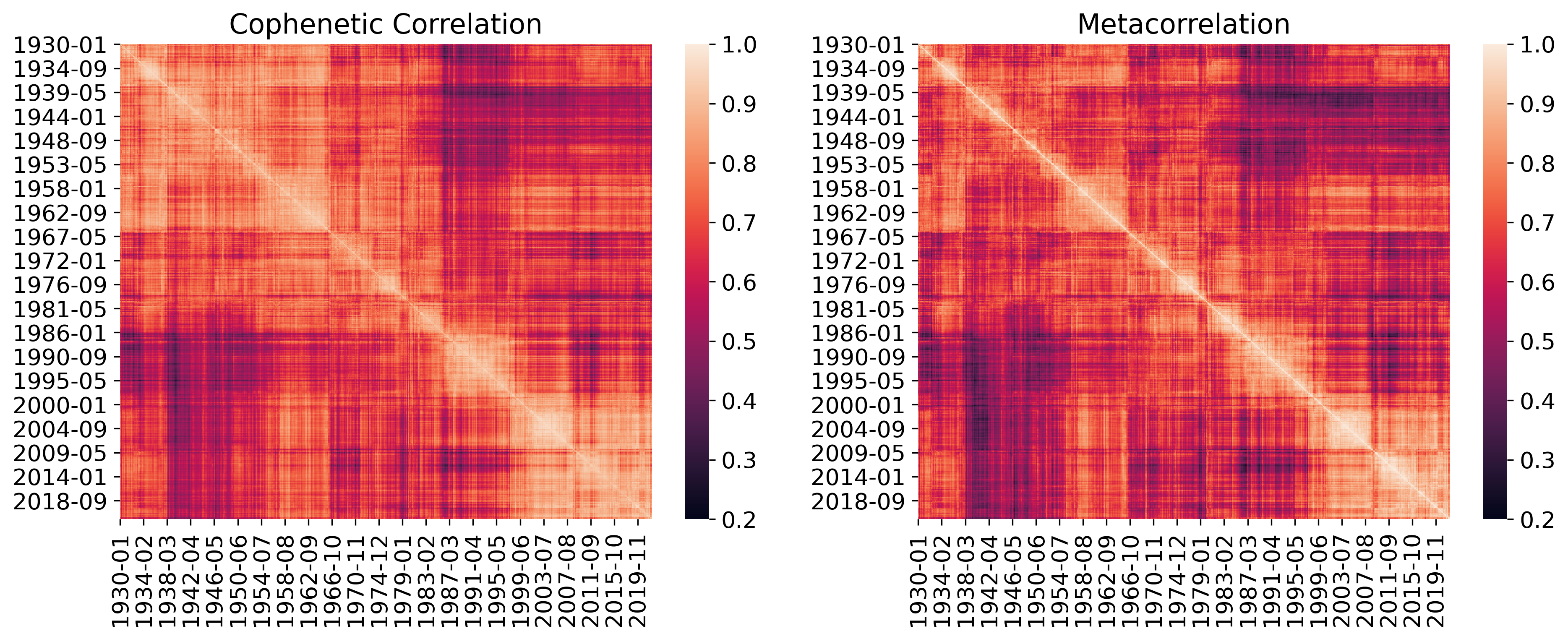}
  \caption{Similarity of moments in time measured via the cophenetic correlation or metacorrelation on matrices of $\tau_K^w$ correlations of returns. Windows long four years are used.}
  \Description{Two heatmaps showing clear block structure.}
  \label{LT-heat-sim}
\end{figure}

\paragraph{Similarity matrix between points in time.} Following the methodology described in Section \ref{method-regimes}, we compute Pearson's, Spearman and (weighted) Kendall correlations on our time series of LT assets' returns using a time window of length four years, which we slide every month. Then, we identify regimes by clustering the matrices of cophenetic correlations or metacorrelations computed on the above sets of correlations.
Figure \ref{LT-heat-sim} shows these matrices of time-similarity from initial weighted Kendall correlations, which is the measure that produces the sharpest patterns. Indeed, the weighting dampens spill-overs of high correlation of furthest points in time in the overall description of each period under consideration. 
The results from other initial correlation options are not shown for the sake of brevity, but we witness general agreement on the clear block structure of each matrix. 
Each mean cophenetic correlation between a dendrogram and its own feature matrix is $>93\%$, meaning that the hierarchical clustering is indeed able to preserve the relevant information of the data. However, the metacorrelation 
matrix provides a slightly stronger diversification in the level of similarity between different points in time, and will be the method of choice in our analyses.

\begin{figure*}[h]
  \centering
  \includegraphics[width=0.8\linewidth]{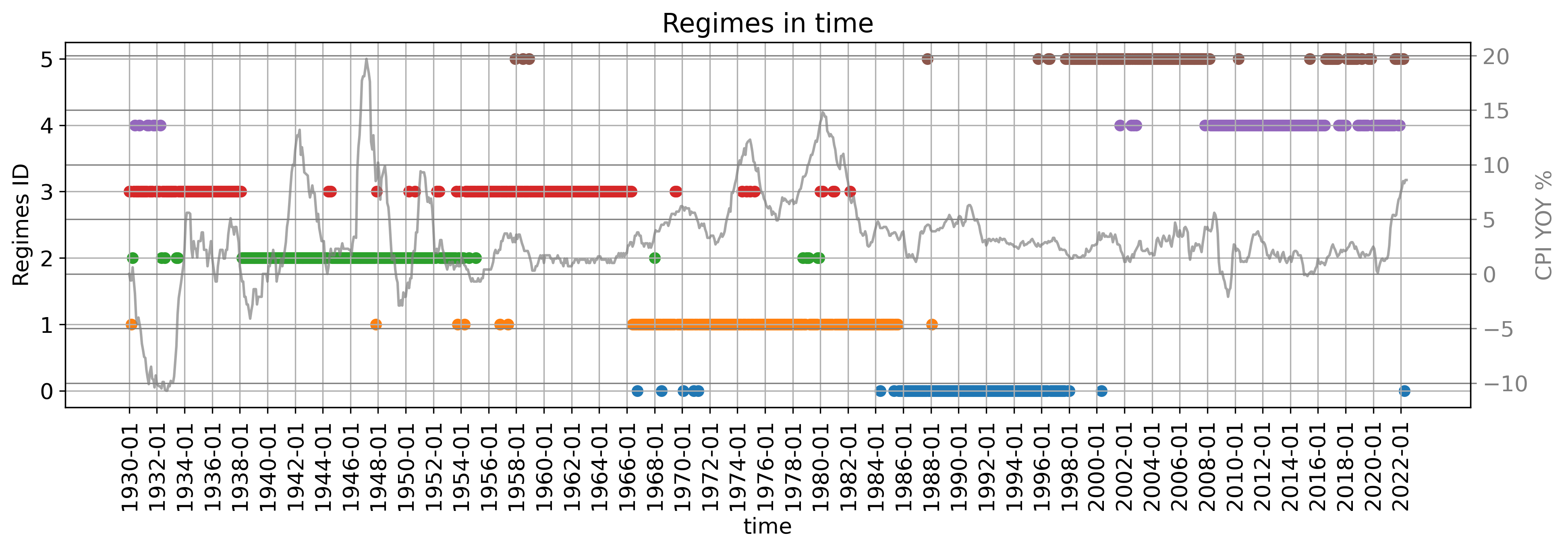}
  \caption{Division of time into six different regimes identified from the correlation of returns of our LT indices. Inflation (CPI YOY $\%$) is also shown as a macroeconomic indicator for comparison.}
  \Description{A plot summarising the identified regimes in time and also showing CPI YOY in percentage.}
  \label{LT-regimes-time}
\end{figure*}

\begin{figure}[h]
  \centering
  \includegraphics[width=\linewidth]{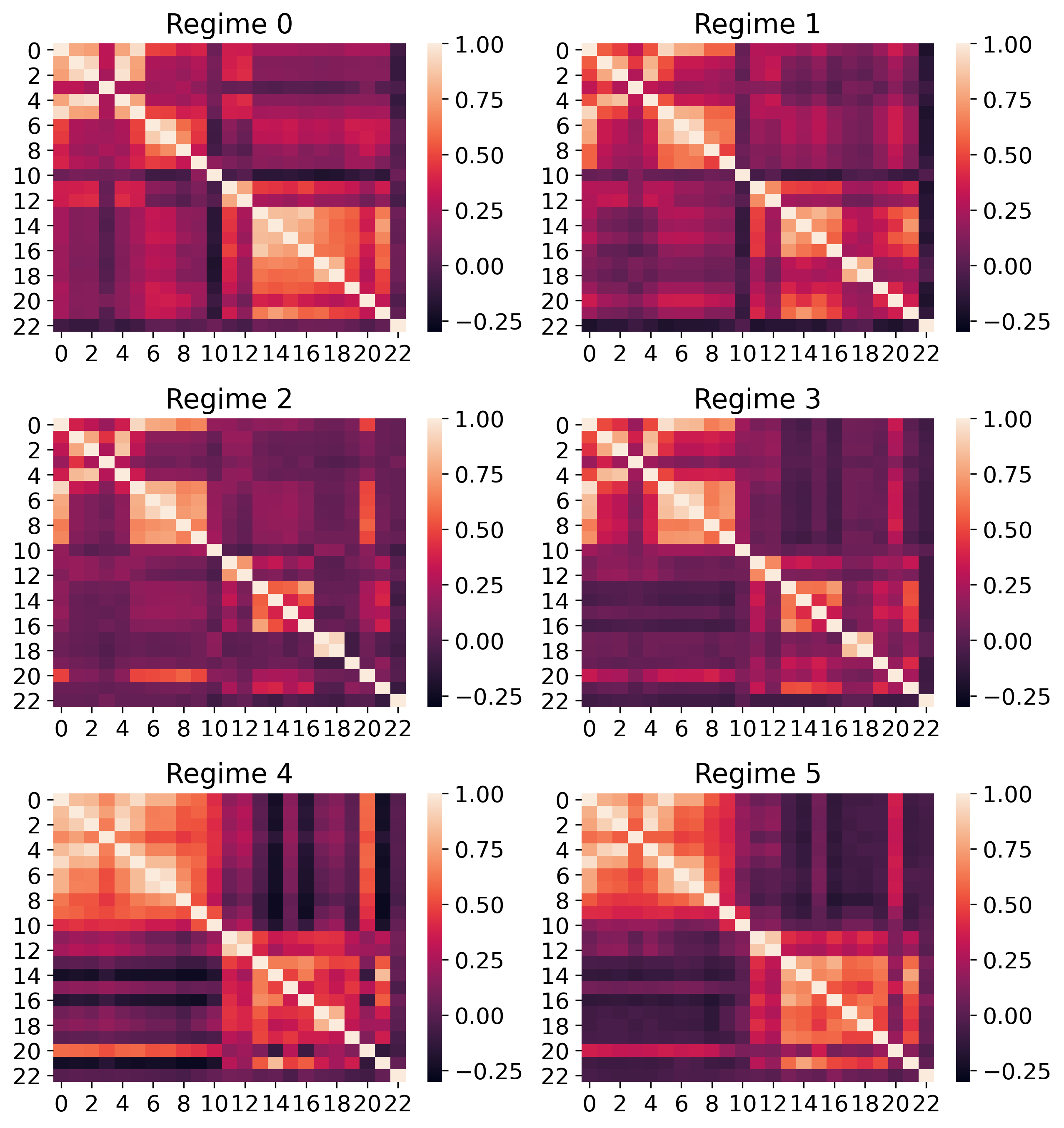}
  \caption{For each regime identified from the LT dataset, we show its matrix of average correlations between indices.}
  \Description{All 6 heatmaps that characterise average correlations between assets in the different regimes.}
  \label{LT-heat-regimes-characterisation}
\end{figure}

\paragraph{Clustering of the similarity matrices.} The time-similarity matrices of Fig. \ref{LT-heat-sim} show a clear block structure that already hints to a related division of periods into regimes. 
We perform PCA on these matrices and find that three dimensions explain $92\%$ of the variance within the data for the cophenetic case, while four dimensions account for $91\%$ of it for metacorrelations.
Next, we cluster points projected onto each new space via KMeans++. The best number of regimes is chosen following the elbow method on an inertia plot, but the stability of results is checked against perturbations of the amount of desired clusters. We also confirm the overall stability of our results by varying initial random seed, time window length and number of dimensions kept in PCA. Our results suggest the existence of \textbf{six} regimes, shown in Fig. \ref{LT-regimes-time} from the metacorrelation matrix. We also plot inflation as Consumer Price Index Year-Over-Year in percentages (CPI YOY $\%$) to have a macroeconomic indicator for comparison. Finally, we calculate the average correlation matrix between assets for each different regime. The results are shown in Fig. \ref{LT-heat-regimes-characterisation}, while Table \ref{LT-table-regimes} reports summary statistics on the related correlations.

\begin{table}[h]
  \caption{Average (absolute) correlation between indices during the six different regimes identified from the LT dataset}
  \label{LT-table-regimes}
  \begin{tabular}{lcccccc}
    \toprule
    Regimes & 0 & 1 & 2 & 3 & 4 & 5\\
    \midrule
    Avg. corr & 0.38 & 0.34 & 0.27 & 0.28 & 0.36 & 0.32\\
    Avg. |corr| & 0.43 & 0.41 & 0.32 & 0.35 & 0.47 & 0.42\\
  \bottomrule
\end{tabular}
\end{table}

\paragraph{Discussion.} Table \ref{LT-table-regimes} highlights how regimes 
R$2 \; \& \; \text{R} 3$ have much lower average correlations between indices' returns than the others. These are quieter regimes that span years from the $'30$s to mid $'60$s, as depicted in Fig. \ref{LT-regimes-time}. While they encapsulate World War II, R$2 \; \& \; \text{R} 3$ can be seen as representatives of the Golden Age associated with the post-War economic boom. Regime $3$ ends in the mid $'60$s, when indeed economists mark the beginning of the Great Inflation (from $1965$ to $1982$, \url{https://www.federalreservehistory.org/}). The Great Inflation can be recovered in our R$1$ and is followed by a moment of recession. The latter relates to R$0$, which indeed has higher average correlations. However, the highest average correlations happen in R$4$ when both the Great Financial Crisis (GFC) and COVID-19 hit.

To delve deeper into the identified regimes, we focus on their characteristic heatmaps of Fig. \ref{LT-heat-regimes-characterisation} that show the average correlations between indices' returns. As expected, R$4$ has the strongest block structure that underlines broadly high correlations within equity indices (IDs $0$-$9$) and within fixed income ones (IDs $11$-$21$). Furthermore, there are also significant off-diagonal correlations especially on the negative range. Regime $5$ shows a similar structure but less enhanced, completing the period of the $2000$s. We could propose to agglomerate our six regimes into three major ones ($3\&2$, $1\&0$ and $5\&4$) by looking at the heatmaps' structure. However, the importance of the full discretisation becomes clear once we compare our set of regimes with the CPI YOY $\%$ index. The levels of inflation and inflation's volatility vary substantially over time and can be associated to the regimes identified.

Moving to specific indices, we can see how commodity (ID $10$) has neutral or slightly negative correlations to equity in R$1 \; \& \; \text{R} 0$ but the two categories become strongly positively correlated especially in R$4$. Then, IDs $14\&16$ (risk-free, US Treasury investments) and ID $20$ (higher risk, below investment grade rated investments) strongly react to market distress as intuitively expected. They show negative correlation in R$4$ but a positive relation during the Great Inflation.

Overall, we are able to extract data-driven regimes that are both intuitive and significant from a macroeconomic perspective. Thus, we proceed to deploy our methodology also on the BBG dataset.



\begin{table}[h]
  \caption{Average (absolute) correlation between indices during the seven regimes identified from the BBG dataset}
  \label{BBG-table-regimes}
  \begin{tabular}{lccccccc}
    \toprule
    Regimes & 0 & 1 & 2 & 3 & 4 & 5 & 6\\
    \midrule
    Avg. corr. & 0.06 & 0.05 & 0.04 & 0.08 & 0.06 & 0.05 & 0.08\\
    Avg. |corr.| & 0.15 & 0.14 & 0.13 & 0.18 & 0.15 & 0.14 & 0.16\\
  \bottomrule
\end{tabular}
\end{table}

\begin{figure*}[h]
  \centering
  \includegraphics[width=0.8\linewidth]{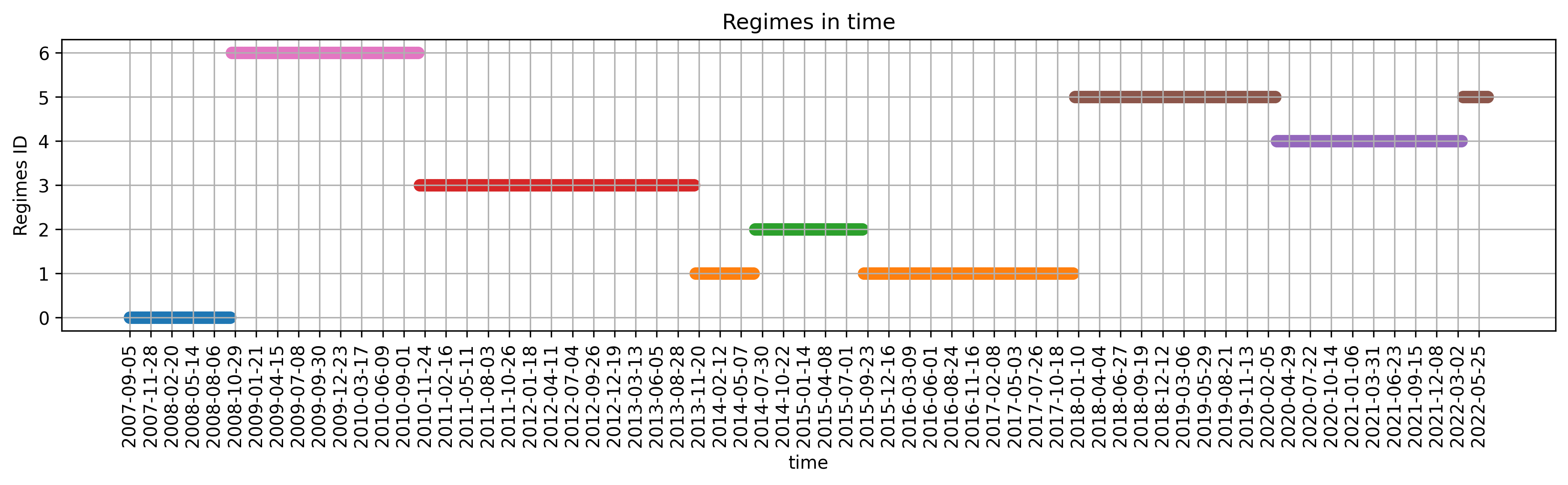}
  \caption{Division of time into seven different regimes identified from the correlation of returns of our BBG indices.}
  \Description{A plot summarising the identified regimes in time for the BBG dataset.}
  \label{BBGregimes-time}
\end{figure*}

\begin{figure*}[h]
  \centering
  \includegraphics[width=0.9\linewidth]{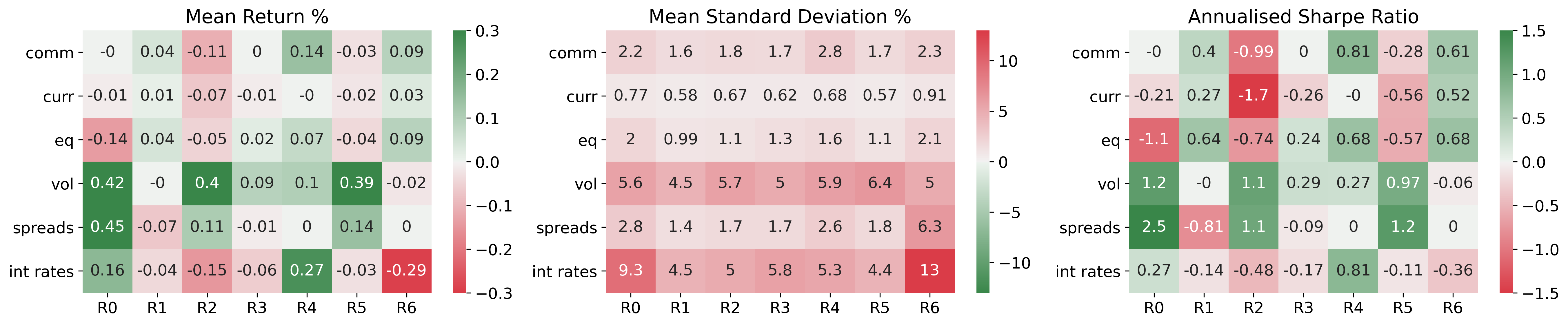}
  \caption{Mean return, standard deviation and 
  annualised Sharpe Ratio of the indices belonging to each asset class divided by regime for the BBG dataset. Note that this akin to a long-only portfolio containing indices from a specific asset class. 
  } 
  \Description{Three plots for the relative characteristics of asset classes' investments during different regimes.}
  \label{BBG-division-ret-std}
\end{figure*}

\subsection{BBG data set}

\paragraph{Identification of regimes.} We compute correlations between BBG indices' returns using a window of two years length that slides one week at every iteration. Motivated by similar considerations to the ones for the LT data set, we focus on the extraction of regimes from metacorrelations between flattened matrices of weighted Kendall correlations. The matrix of periods' similarities shows again a clear block structure but is not reported here for the sake of brevity, as other following intermediate steps.
We do PCA and see that three dimensions describe $\sim 90\%$ of the variance of the data. Points in time are projected onto this new space and clustered via KMeans, where the optimal number of groups is again chosen by looking at the inertia loss function. We find an optimal discretisation of time into \textbf{seven} regimes, which are shown in Fig. \ref{BBGregimes-time}.
Average (absolute) correlations are reported in Table \ref{BBG-table-regimes}, while Fig. \ref{BBG-division-ret-std} highlights the average mean return, average standard deviation and Sharpe Ratios among asset classes during each regime.

Interestingly, R$3$ is the regime with highest absolute correlation and relates to the period between the end of $2010$ and end of $2013$.
In $2010$, there was the first drop in oil prices otherwise rallying in the related commodity super-cycle. However, this is also one of the moments of strong rounds of Fed's Quantitative Easing. By looking at the variation of standard deviation in interest rates' returns, we can indeed see how this quantity significantly dropped in regimes after the end of $2010$. Our intuition is that the inflow of liquidity might have destabilised correlation levels when considering the broader system of all asset classes.


\begin{figure*}[h]
\centering
\includegraphics[width=0.8\linewidth]{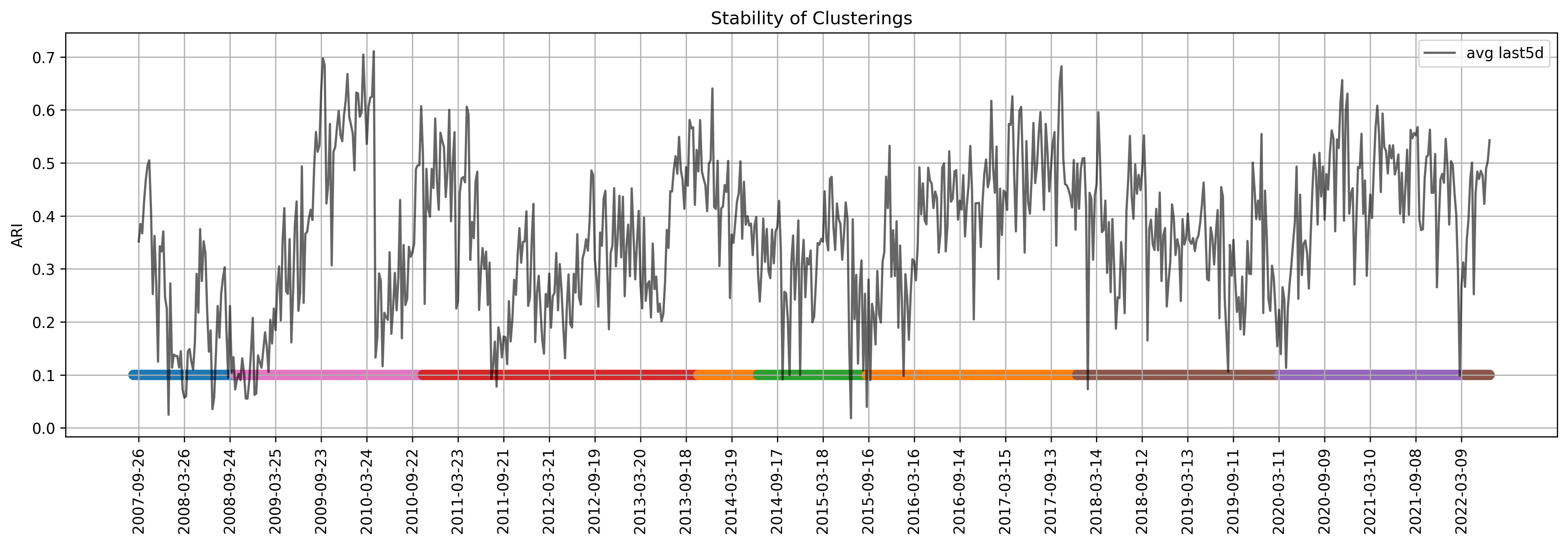}
\caption{Stability of communities detected in time. We show the average ARI between clusterings from the latest five dates and find agreement between drops in the measure and regime changes identified from the metacorrelation matrix.}
\Description{Stability of network's clusters in time from an average ARI value over the past month and previously identified regimes from metacorrelations.}
\label{BBG-ARI-regimes}
\end{figure*}

\begin{table*}
  \caption{Nodes with highest betweenness centrality for each regime identified from the BBG dataset}
  \label{BBG-table-betweenness}
  \begin{tabular}{cccccccc}
    \toprule
     Regimes & 0 & 1 & 2 & 3 & 4 & 5 & 6\\
     \hline
     Rank &  &  &  &  &  &  & \\
    \midrule
    $1^{st}$ & MSCI US& MSCI EAFE & US Corp10-25Yr & MSCI Europe & SGDUSD & MSCI Europe & US Corp5-10Yr\\
 \hline
 $2^{nd}$ & US Corp5-10Yr & DE 10Y & DE 10Y & AUDUSD & MSCI Europe & AUDUSD & MSCI Pacific\\
 \hline
 $3^{rd}$ & BBG COMM & MSCI Italy & MSCI Europe & BBG COMM & MSCI EM Asia & AU 10Y & BBG COMM\\
 \hline
 $4^{th}$ & MSCI Europe & SE 10Y & MSCI EAFE & US Corp10-25Yr & MSCI US & US Infor. Tech.& KR 2Y\\
 \hline
 $5^{th}$ & EAFE SmallCap & SZ 10Y & MSCI EM & MSCI US & JP 10Y & DE 10Y & MSCI World\\
    \bottomrule
  \end{tabular}
\end{table*}

\begin{figure}[h]
\centering
\includegraphics[width=0.97\linewidth]{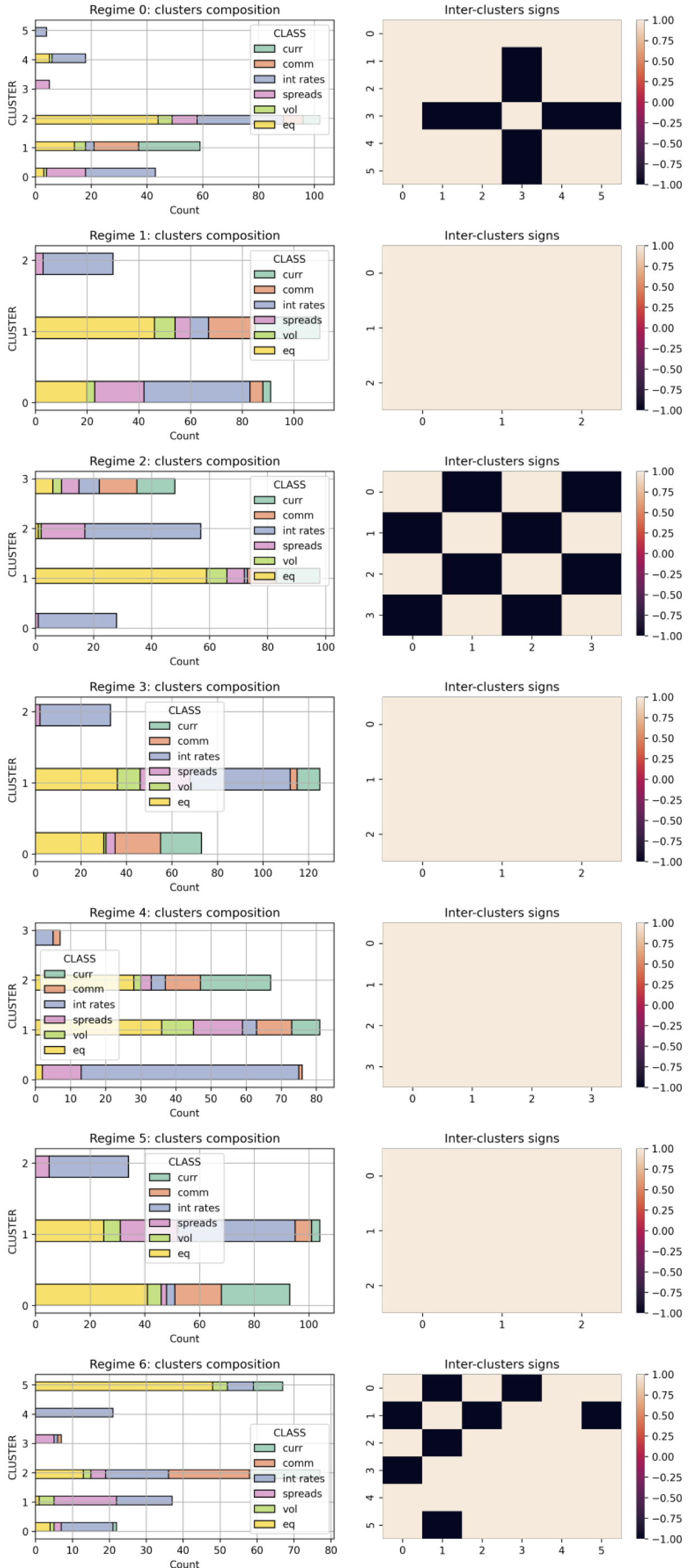}
\caption{For each BBG regime, we plot the composition of communities identified by clustering the related signed network (left column). We also show the sign of average correlations between the different clusters (right column).}
\Description{BBG-characterisation of regimes via inner clusters.}
\label{BBG-categories into clusters}
\end{figure}

\paragraph{Network-based identification of regimes.} We increase the confidence in our regimes by separately studying the structural evolution in time of the network of returns' correlations between indices.

We build one network per week following the adopted sliding window. 
Nodes are the BBG indices and weighted edges are added from the related correlations if their magnitude is above the threshold $|\tau_K^w| \geq 0.2$. Then, the giant component of each network is kept.
A higher number of survival links is an indicator of increasing correlations between assets and points towards periods of market distress. For the interested reader, we mention that this is indeed verified for the evolution of our network when the introduced indicator is compared to peaks in the VIX Volatility Index. However, the decay of volatility levels is faster than the dissipation of links.

We proceed by clustering nodes of each signed network via the symmetric SPONGE algorithm. The similarity between communities extracted at consequent points in time is computed by averaging the ARI value between the current clustering and each clustering for the past 
four weeks. 
The result is plotted in Fig. \ref{BBG-ARI-regimes}, where we also show in different colors the regimes previously found from the metacorrelation matrix.
To choose the number of groups to cluster for at each iteration, we automatically pick the value for which there is the highest increase in 
$Q_{signed}$ 
when clustering for $k \in [2, ..., 10]$. This method is superficial but suffices for building a summary measure to validate our regimes. Indeed, we can see general 
consensus between strong drops in stability of the ARI average and a transition among the previously identified regimes.


\paragraph{Network-based characterisation of regimes.} We begin by calculating seven distance matrices $D_{R}=\sqrt{2 \cdot (1-|\tau_{K,R}^{w,avg}|)}$ from the average correlation matrices of regimes, and build the related graphs. 
We investigate which nodes have the highest betweenness centrality for each regime, and observe that the distribution of centralities varies substantially and reaches the highest magnitudes in regime  R$2$. Table \ref{BBG-table-betweenness} reports the five nodes with the highest betweenness centrality for each regime. Thanks to this setup, we understand the most influential indices on the propagation of trends during different periods. As an example, it is interesting to notice how the Singaporean exchange rate (SGDUSD) is the main bridge for the diffusion of the COVID-19 shock due to its highest centrality in R$4$.

We also directly consider the average correlation matrix within each regime, build the related signed graph, and cluster the nodes via the symmetric SPONGE algorithm. The optimal number of communities $k^R$ for each regime is identified by looking at the signed modularities for $k \in [2, ..., 15]$. The results are depicted in Fig. \ref{BBG-categories into clusters}, where we show the composition among asset classes of each identified community for the different regimes. In parallel, we also report the sign of average correlations between each pair of communities and notice that only three regimes show some dominant negative correlations between clusters. Focusing on interest rates, we know that they can be divided into risk-free ones (e.g. Germany, Switzerland) or riskier options (e.g. Chile, Indonesia). Looking at the histograms for the latest R$4 \; \& \; \text{R}5$ and the full composition of communities, we observe that this division is violated during the COVID-19 crisis. Almost all indices related to interest rates are clustered together in R$4$, meaning that even the safest investments acquired some level of risk.

As intended, our framework sheds further light on the evolution of broad market dynamics, whose understanding 
is essential 
for better investment decision and risk-on/risk-off attitude switching.

\section{Results: Lead-lag clusters}   \label{sec:resLeadLag}

For each regime depicted in Fig. \ref{BBGregimes-time}, we extract the related partial time series of daily returns for all our assets. Then, we compute Granger causalities between indices for different lags in each regime, following the methodology highlighted in Section \ref{sec:ll}. The lag generating the highest number of significant relationships at the $0.05$ level (generally over $50\%$) is kept as characteristic of the regime, and the resultant relationships are encoded in a skew-symmetric matrix. These optimal lags $g^R$'s are reported in Table \ref{BBG-table-returns}, where we can notice that R$5$ has a much longer-than-average $g^5$, of $42$ trading days, and relates to turning points of monetary tightening.

\begin{figure}[h]
  \centering
  \includegraphics[width=0.7\linewidth]{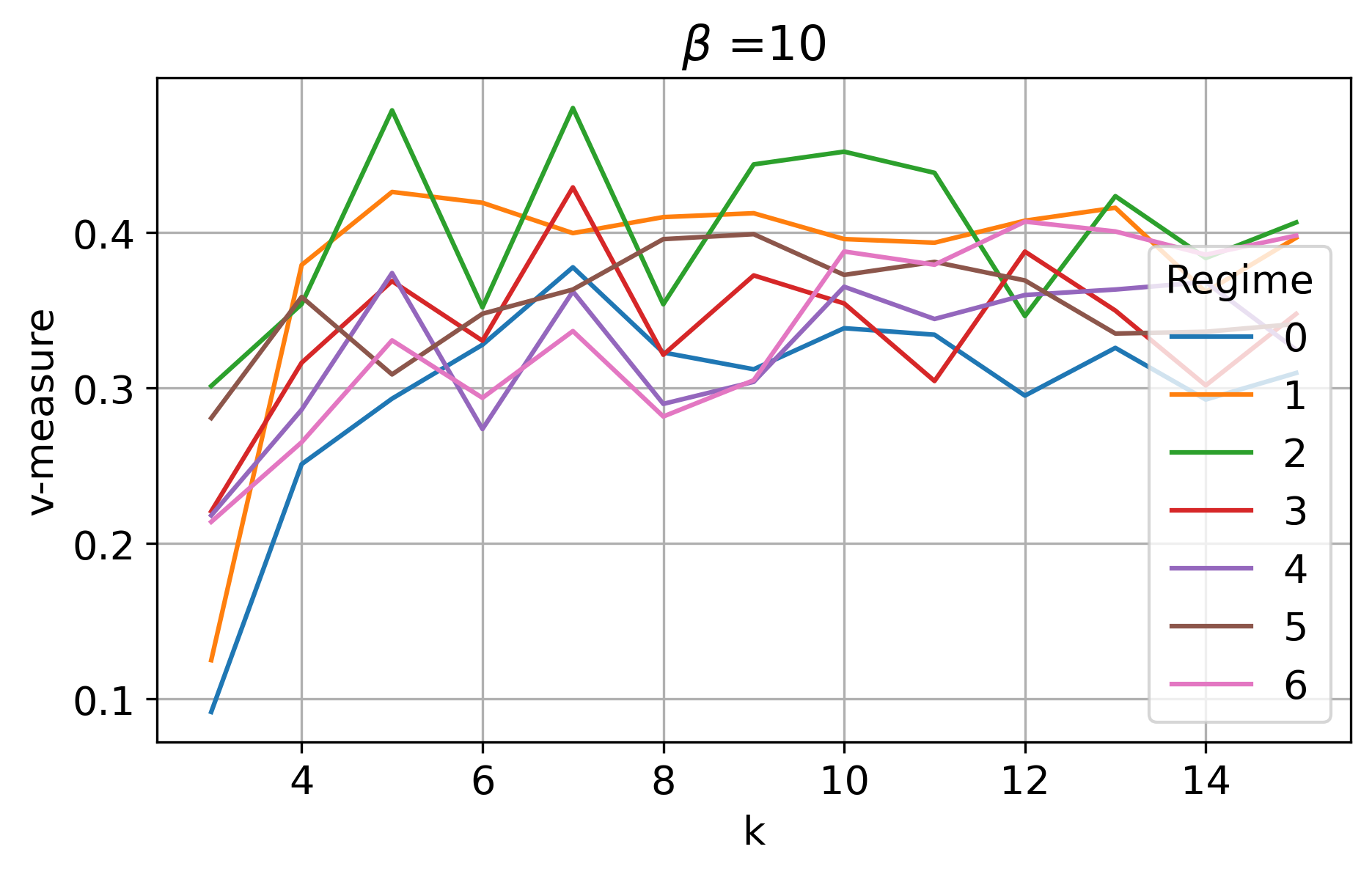}
  \caption{The $v-measure$ is computed as a measure of goodness of the recovered communities in each regime. We vary the number of groups $k$ and favour completeness via $\beta = 10$.}
  \Description{v-measure as goodness.}
  \label{betas}
\end{figure}

We build a directed network from the Granger causality matrix of each regime R. Then, we cluster its nodes with the Hermitian clustering algorithm to understand evolving communities of major leaders and laggers. To define the number of clusters $k^R$ to aim for, we recall that
our indices belong to six asset classes with likely intrinsic structural similarities. 
Therefore, 
we decide to investigate the forecasting power between asset classes allowing for inner divisions.
We test $k \in [2, ..., 15]$, and compute for each clustering the related $v-measure$ with $\beta=10$, as shown in Fig. \ref{betas}. While other values of $\beta$ are checked, we follow this choice to have both homogeneity and completeness of the extracted communities, but favouring the latter. A too granular view would have been 
suggested with lower $\beta$'s, \mc{hindering the interpretability of the results.}  

A first rough jump is generally witnessed for $k=5$ in Fig. \ref{betas}. For simplicity, we directly use this value as the fixed number of communities to look for in all regimes.
We proceed with the clustering and compare both the strength of links between clusters and the clusters' composition within each regime. While the full set of plots is not shown for the sake of brevity, they reveal that the composition of the most leading and most lagging clusters varies significantly between regimes. These are usually equities versus interest rates, but also currencies become meaningful leading assets in R$3 \; \& \; \text{R}6$. 
As a final intuitive example, we mention that the most leading cluster in R$0$ (GFC) is composed of US equities and US volatilities, while European interest rates are the laggers.

To conclude, we compare the average return of a portfolio constructed by considering  the most leading and lagging clusters versus a plain investment on all indices. Table \ref{BBG-table-returns} shows our results. In each regime, we have positive return (reported in percentage) that also outperform the benchmark, apart for R$4$. While this is not a real investment strategy, the framework shows the importance of recognising both regimes and their inner evolving leaders and laggers, \mc{thus providing insights towards a better portfolio construction}.

\begin{table}
  \caption{Average return of an investment based on lead-lag clusters or on the full universe of indices for each regime. The optimal lag $g^R$ of each regime is also reported}
  \label{BBG-table-returns}
  \begin{tabular}{lccccccc}
    \toprule
    Regimes & 0 & 1 & 2 & 3 & 4 & 5 & 6\\
     \hline
     $g^R$ & 10 & 1 & 1 & 1 & 5 & 42 & 5\\
    \midrule
    Lead-Lag avg. ret $\%$ & 5 & 1 & 0.1 & 0.8 & 0.2 & 0.3 & 0.5\\
 Benchmark avg. ret $\%$ & 1 & -0 & -0 & -0 & 0.6 & 0.2 & -0.3\\
  \bottomrule
\end{tabular}
\end{table}



\section{Conclusions}   \label{sec:concl}

The aim of this work is to shed more light on the co-evolution of relationships between asset classes and varying financial market  conditions. We are able to identify macro regimes by clustering periods in time that show similar correlations of returns of indices belonging to different asset classes.
Then, we develop an initial investment framework that leverages on regime-specific groups of securities that lead and lag each other, in order to optimise asset allocation. 
Our work highlights the importance of regimes' characterisation and regime-transition identification for investment decisions. We lay the foundations for a measure of stability of regimes that could suggest moments in which to adopt a risk-off attitude, and provide a first purely data-driven characterisation of regimes that can be used for reference with future market developments. As next steps, our efforts will focus on stronger lead-lag strategy development and out-of-sample testing \mc{of the uncovered regimes and relationships}.



\begin{acks}
Deborah Miori's research was supported by the \textit{EPSRC CDT in Mathematics of Random Systems} (EPSRC Grant  EP/S023925/1).
\end{acks}

\bibliographystyle{ACM-Reference-Format}
\bibliography{zz_main}

\appendix

\section{Lists of Indices}

\subsection{LT dataset}
\label{A-LT-data}

The full list of indices with related (ID) is here reported for the LT dataset. The indices are divided into $4$ asset classes, namely \textit{equity, commodity, fixed income} and \textit{cash}. 

\paragraph{Equity.} ACWI-MSCI All Country World (0), ACWIxUS-MSCI All Country World ex-US (1), EAFE-MSCI EAFE (2), EM-MSCI Emerging Markets (3), DM-MSCI World ex-US (4), World-MSCI World (5), S\&P 500 (6), EQ-Dow Jones US Total Stock Market (7), Russell 2000 (8), NREQ-S\&P North Am. Natural Resources Sect. (9).

\paragraph{Commodity.} COMM - BBG Commodity (10).

\paragraph{Fixed income.} BBG Barcalys: Global Treas. (11), Global Treas. ex-US (12), IG- US Aggr. Bond (13), US Long Treas. (14), US Long IG Corp. Bonds (15), IT- US Interm. Treas. Bond (16), TIP- US TIPS (17), US 1-10Y TIPS (18), MUNI- Municipal Bond (19). Then, HY-ICE BofAML US HY Constr. (20) and BBG US STRIPS 25-30Y (21).

\paragraph{Cash.} BBG Barclays US 1-3 Months Treasury Bills (22).


\subsection{BBG dataset}
\label{A-BBG-data}

The full list of indices considered in the Bloomberg BBG dataset follows. These are divided into $6$ asset classes, namely \textit{commodity, currency, equity, volatility, bond spreads} and \textit{interest rates}.

\paragraph{Commodity.} BBG Commodity Index. Futures $'22$:
    GOLD 100 OZ  Jun, 
    NATURAL GAS  Jul, 
    BRENT CRUDE  Aug, 
    WTI CRUDE  Jul, 
    SOYBEAN    Jul, 
    CORN       Jul, 
    COPPER     Jul, 
    Low Su Gasoil G   Jun, 
    SILVER     Jul, 
    LIVE CATTLE  Jun, 
    SOYBEAN OIL  Jul, 
    WHEAT (CBT) Jul, 
    LME PRI ALUM Jun, 
    LME NICKEL Jun, 
    NY Harb ULSD  Jul, 
    SOYBEAN MEAL Jul, 
    LME ZINC   Jun, 
    SUGAR \#11 (WORLD) Jul,
    COFFEE C Jul,
    KC HRW WHEAT  Jul, 
    LEAN HOGS  Jun, 
    COTTON NO.2  Jul.

\paragraph{Currency (-USD)} AUD, 
BRL, 
CAD, 
CHF, 
CLP, 
CNY, 
COP, 
CZK, 
EUR, 
GBP, 
HUF, 
IDR, 
INR, 
JPY, 
KRW, 
MXN, 
MYR, 
NOK, 
NZD, 
PEN, 
PHP, 
PLN, 
RUB, 
SEK, 
SGD, 
THB, 
TRY, 
ZAR.

\paragraph{Equity (MSCI Index in USD)} World, 
North America, 
EAFE, 
Europe, 
Nordic, 
Pacific, 
Far East, 
Canada, 
USA, 
Austria, 
Belgium, 
Denmark, 
Finland, 
France, 
Germany, 
Israel, 
Ireland, 
Italy, 
Netherlands, 
Norway, 
Portugal, 
Spain, 
Sweden, 
Switzerland, 
UK, 
Australia, 
Hong Kong, 
Japan, 
New Zealand, 
Singapore, 
EM Europe, 
EM Latin Am, 
EM Asia,
North America Small Cap, 
EAFE Small Cap, 
Pacific Small Cap.
For each one of Europe, USA and EM: Energy, Materials, Industrials, Consumer Discret., Consumer Staples, Health Care, 
Financials, Infor. Tech., Comm. Services, Util. Sector.

\paragraph{Volatility.} Cboe, 
Cboe NDX, 
Cboe RSL2000, 
VSTOXX, 
MOVE, 
NIKKEI, 
HSI, 
JPM JP, 
Cboe20+Y TreasBnd, 
JPM G7, 
JPM EM.

\paragraph{Bond Spreads.} US Corp: 1-5Y, 5-10Y, 10-25Y, 25+Y. 
Pan-Euro Corp.
Pan-Euro Corp: 1-3Y, 3-5Y, 5-7Y, 7-10Y, 10+Y.
Global Agg Index, 
US IG Corp,
US HY Corp, 
US Gov-Related, 
US ABS, 
US CMBS, 
US MBS, 
CAD Credit, 
EuroAgg Gov-Related, 
EuroAgg Securitized, 
EuroAgg High Yield.
Sterling Agg: Corp, 
Sterling Aggregate: Gov-Related, 
Japan Corp, 
Australia Corp, 
China Corp, 
EM USD IG, 
EM USD HY.

\paragraph{Int. Rates (2,10,30Y)} AU,
CA, 
CL, 
CN, 
CZ, 
DE, 
DK, 
FI, 
GB, 
HK, 
HU, 
ID, 
JP, 
KR, 
MX, 
MY, 
NO, 
NZ, 
PL, 
SE, 
SG, 
SZ, 
TH, 
US CMT, 
ZA.

\end{document}